\documentclass[a4paper, 10pt, conference]{ieeeconf}      % Use this line for a4 paper

\IEEEoverridecommandlockouts                              % This command is only needed if 
\usepackage{graphics} % for pdf, bitmapped graphics files
\usepackage{epsfig} % for postscript graphics files
\usepackage{amsmath} % assumes amsmath package installed
\usepackage{amssymb}  % assumes amsmath package installed

\title{\LARGE \bf Stochastic Model Predictive Control and Sewer Networks}

\author{Jan Lorenz Svensen$^{1}$, Hans Henrik Nieman$^{2}$, Anne Katrine Vinther Falk$^{3}$ and Niels Kj\o lstad Poulsen$^{4}$% <-this % stops a space
\thanks{*This study was done as part of the Water Smart City Project fonded by Innovation Fund Denmark, as project 5157-00009B}% <-this % stops a space
\thanks{$^{1}$Jan Lorenz Svensen is with the Department of Applied Mathematics and Computer Science,
        Technical University of Denmark, 2800 Kgs. Lyngby, Denmark
        {\tt\small jlsv@dtu.dk}}
\thanks{$^{2}$Hans Henrik Niemann with the Department of Electrical Engineering, Technical University of Denmark,
        2800 Kgs. Lyngby, Denmark
        {\tt\small hhn@elektro.dtu.dk}}
\thanks{$^{3}$Anne Katrine Vinther Falk with the DHI Denmark,        2970 H\o rsholm, Denmark
        {\tt\small akf@dhigroup.com}}
\thanks{$^{4}$Niels Kj\o lstad Poulsen is with the Department of Applied Mathematics and Computer Science,
        Technical University of Denmark, 2800 Kgs. Lyngby, Denmark
        {\tt\small nkpo@dtu.dk}}
}

\begin{document}

\maketitle
\thispagestyle{empty}
\pagestyle{empty}

%%%%%%%%%%%%%%%%%%%%%%%%%%%%%%%%%%%%%%%%%%%%%%%%%%%%%%%%%%%%%%%%%%%%%%%%%%%%%%%%
\begin{abstract}
In this work, an evaluation of Chance-Constrained Model Predictive Control (CC-MPC) in sewer systems over the use of the classical deterministic Model Predictive Control (MPC) is presented. 
The focus of this evaluation is on the avoidance of weir overflow when uncertainty is present. Furthermore, the design formulation of CC-MPC is presented with a comparison to the design of MPC. For the evaluation, a simplified model of the Barcelona sewer network case study is utilized. Our comparison shows that for sewer systems with uncertain inflows, a CC-MPC allows for better statistical guarantees for avoiding weir overflow, than relying on a deterministic MPC. A simple back-up strategy in case of infeasible optimization program was also apparent for the CC-MPC based on the results of the analysis. 
\end{abstract}

%%%%%%%%%%%%%%%%%%%%%%%%%%%%%%%%%%%%%%%%%%%%%%%%%%%%%%%%%%%%%%%%%%%%%%%%%%%%%%%%
\section{INTRODUCTION}
For the last few decades, the usage of Model Predictive Control (MPC) in sewer systems has been researched\cite{GRJ}-\nocite{OMB,Nadia18,QBJ,GMJ}\cite{MPB}. 
While the previous research has primarily focused on deterministic scenarios, known systems and inflows, we will in this work focus on the uncertainty there in reality does exist in sewer systems and how it can be handled. 
Some of the uncertainties there exist with regard to controlling and observing sewer systems includes model and inflow deviations from the expected. 
The uncertainty of the inflow is especially important for the MPC, given its reliance on the predicted forecast of the inflow. Therefore the type of uncertainty considered in this work will be limited to the uncertainty of the inflow. 

While the deterministic MPC previously considered in research, is not directly designed to handle uncertainties in the optimization, stochastic formulations of MPC do exist\cite{Mesbah2016}-\nocite{Grosso2014, ECK1,CGP09}\cite{KC16}.  These stochastic MPCs are designed to handle uncertainties to some degree, and by different approaches depending on the type of stochastic MPC chosen. 
These approaches have wide ranges; some approaches might consider a statistic-based robust method through finite realizations of the uncertainties\cite{CGP09}, while others might consider expectation or probabilistic constraints to handle the uncertainties\cite{KC16}, some with additive uncertainties\cite{Grosso2014} others with multiplicative uncertanties\cite{ECK1}. In this work, we investigate a probabilistic method known as Chance-Constrained MPC (CC-MPC), previously applied to drinking water networks\cite{Grosso2014}.

\begin{figure}
    \centering
     \includegraphics[width = 0.4\textwidth,trim={2.6cm 20.6cm 9.1cm 0cm},clip ]{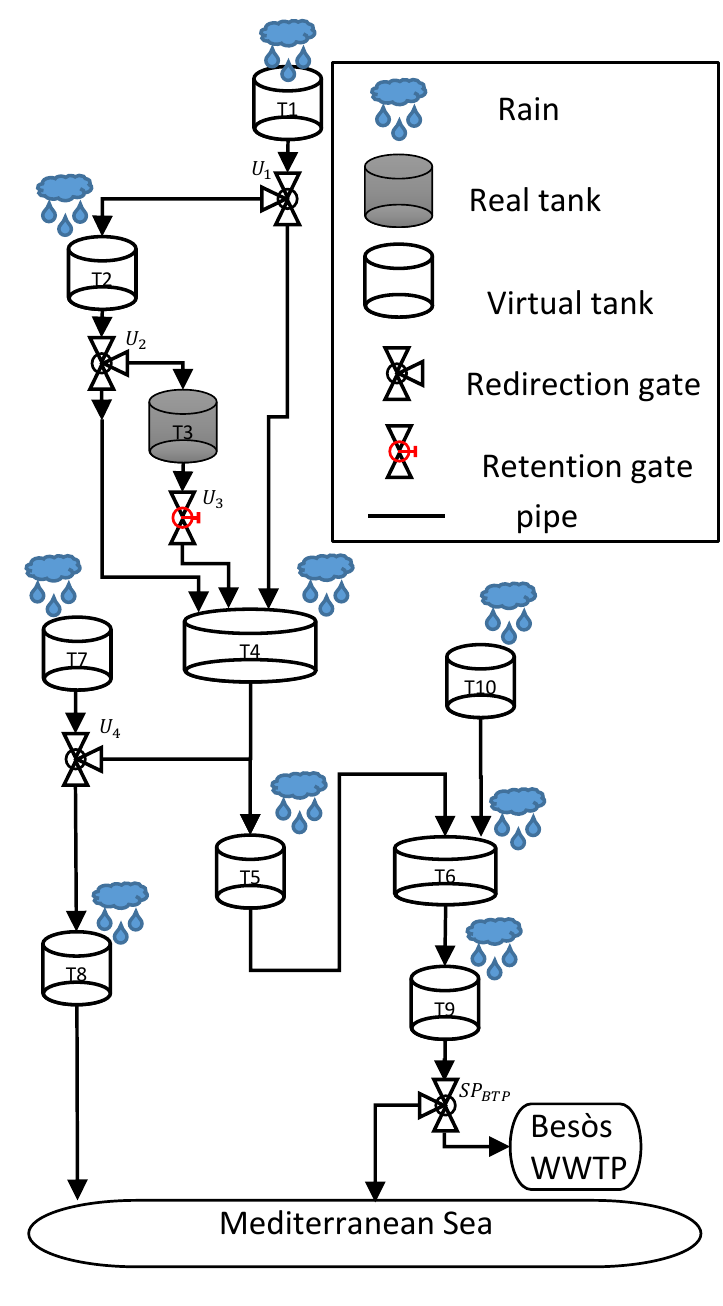}
     \caption{A schematic of the simplified model inspired by the Barcelona Sewage System model presented by Ocampo-Martinez \cite{OMB}, showing the interconnections between the different tanks and the environment.  The controlled parts of the system are tagged with a U}
      \label{fig:Barcelonal}
\end{figure}

For an evaluation and comparison of  both the deterministic and the stochastic MPCs' performance, the results of simulations are utilized. The evaluation will primarily focus on their performance in avoiding weir overflows from occurring. In the evaluations, a model inspired by the Barcelona Case study\cite{OMB} is utilized. The model, seen in Fig. \ref{fig:Barcelonal}, is a simplified version of its inspiration formulated as a linear model.
\subsection{Notation}
In this paper, the following notations are employed. Bold font is used to write vectors, while the quadratic norm of $\textbf{x}$ is written as $||\textbf{x}||^2_A=\textbf{x}^TA\textbf{x}$, while a generic set of function varibables are indicated by the bullet $\bullet$.
The functions $E(x)$,  $Pr(X\leq x)$ and $\Phi_F(x)$ indicate the expectation of x, the probability of X less than x, and the cumulative density function(CDF) of a given distribution F respectively. 
The notation $X\sim F$ indicates that X is following a given distribution F, While distributions written as $F_a^b$ indicate bounds on the distribution, a for the lower and b for the upper bound. Distributions written without parameters are standardized distributions, e.q. normal distribution with zero mean and unit variance.
The variables written as $\mu_x$ and $\sigma^2_x$ are the mean and variance of a variable x. The min. and max. of a function $f(x)$ is denoted by $\underline f$ and $\bar f$ respectively.
The superscripts indicate the nature of the parameter with the following meanings: u means a controlled variable, w means it relates to a weir, $ref$ means it is a reference, and $in$ and $out$ denotes inflow and outflow respectively. The subscript k indicates the sample time, while the sampling time of the system is written as $\Delta T$.  A variable noted by a V or q is representing a volume or a flow respectively.

\section{Models}
Two models are utilized for the evaluation, one for the simulation and one for the control design. The models are based on the Barcelona sewage system model\cite{OMB} but reduced such that the system only contains ten tanks, and only has weirs in the virtual tank elements. Given that the aim for the controllers is to avoid weir overflows, the weirs are not included in the model for the MPC. The simplified models can be described as a combination of a few components, assembled as in Fig. \ref{fig:Barcelonal}. The main component is the virtual tank, given by (\ref{eq:VT1}) and (\ref{eq:VT2}).  For the simulation model, the weir overflow is given by (\ref{eq:VT3}), while it is zero in the control model. The control model also includes boundaries on the tank volume given in (\ref{eq:VT4})
\begin{align}
	V_{k+1} =& V_{k} + \Delta T (q^{in}_k-q^{out}_k - q^w_k) \label{eq:VT1}\\
	q^{out}_k =& \beta (V_k-\Delta T q^w_k)\label{eq:VT2}\\
	q^w_k =& \max\bigg(0,\frac{V_k - \bar V}{\Delta T}\bigg)\label{eq:VT3}\\
	0 \leq& V_k \leq \bar V\label{eq:VT4}
\end{align}
Where $\beta$ is the volume/flow cofficient\cite{Singh1988} of the tank outflow.
The system also contains a real tank, where the difference from the virtual tank is that the weir overflow $q^w_k$ is zero for both models. Another difference is that the outflow $q^{out}$ is controlled by a retention gate $q^{u,out}$. In the control model, the controlled outflow is constrained by:
\begin{align}
	0 \leq q^{u,out} \leq \bar q^{u,out}\\
	q^{u,out} \leq \beta V_k
\end{align}
The last component is the redirection gate, which is given by (\ref{eq:rg1}). The control flow $q^u$ of the redirection gate reduces the outflow and is constrained in the control model by (\ref{eq:rg2})-(\ref{eq:rg3}).
\begin{align}
	q^{out}_k = q^{in}_k - q^u_k \label{eq:rg1}\\
	0 \leq q^u_k\leq q^{in}_k\label{eq:rg2}\\
		q^u_k\leq \bar q^{u}\label{eq:rg3}
\end{align}

\section{MPC design}
The classical MPC is a deterministic controller and is designed based on the assumption of the ideal scenario; where we would have perfect knowledge about the rain inflow to the sewer systems. In MPC, an optimization program is solved at each time instance k in conjuction with the prediction horizon receeding.
The formulation of the optimization program consists of two parts; a cost function and a constraint set, which together form the program. In this paper, we will formulate the program of the MPC as a quadratic program of the minimization variant.  A quadratic program means the terms of the cost function can only be quadratic or linear, and that the constraints are all linear. The cost function utilized in this work is given by (\ref{eq:MPC}) and aims at minimizing the cost terms.
\begin{equation}\label{eq:MPC}
J = \min\limits_{u}\sum\limits_{k=0}^N ||(\textbf{z}_k-\textbf{z}_k^{ref})||^2_Q+||\Delta \textbf{q}^u_k||^2_R
\end{equation}
The terms of the cost functions consist of a penalty on the change of controlled flow $\Delta \textbf{q}^u_k$ for each time step k, similarly, it contains a penalty for deviations of the output $\textbf{z}_k$ from the desired output reference $\textbf{z}_k^{ref}$. The output vector $z_k$ consists of two elements, corresponding to the objectives given below:
\begin{itemize}
\item maximize flow to the Bes\'{o}s Wastewater Treatment Plant
\item minimize the aggregated flow to the Mediterranean Sea
\end{itemize}
The priority of the different terms in the cost function was mostly put on the flow to the sea followed by the flow to the treatment plant, with least priority given to the change in control flow.
The weight matrices $Q$ and $R$ were on this basis chosen to be diagonal matrices, with the $Q$ matrix having the weights of each objective being $0.5$ and $1.0$ respectively, while the $R$ matrix was chosen to $0.01$ uniformly in the diagonal.

The constraint set of the MPC is obtained by combining each component of the control model according to Fig. \ref{fig:Barcelonal}, resulting in a constraint set on the form given below for each time step k.
\begin{align} 
\textbf{V}_{k+1} &= A\textbf{V}_k + B\textbf{q}^u_k +G\textbf{q}^{rain}_k\label{eq:MPCCon}\\
\textbf{z}_k &=  C\textbf{V}_k + D\textbf{q}^u_k + F\textbf{q}^{rain}_k\\
\Delta \textbf{q}^u_k & =\textbf{q}^u_k - \textbf{q}^u_{k-1}\label{eq:MPC0}\\
 M\textbf{q}^u_k&+  P\textbf{V}_k+ S\textbf{q}^{rain}_k \leq \textbf{K}\label{eq:MPC1}
\end{align}
Which consists of the three linear equality constraints in (\ref{eq:MPCCon})-(\ref{eq:MPC0}) represents the tank volume process equation, the output equation, and the change of control flow definition respectively. The inequality constraint in (\ref{eq:MPC1}) describes the boundaries of the specific variables and combinations of them. The variable $\textbf{q}^{rain}_k$ is the external inflows into the sewer system, such as rain and dry weather flows.

\subsection{Stochastic MPC - Chance Constrained MPC}
In stochastic MPC, the system is not assumed to be deterministic as before. Instead are knowledge about the uncertainty of the inflow taken into account during the optimization.
The usage of the knowledge of uncertainty requires that the MPC formulation is changed to accommodate the information. While there exist many different approaches to stochastic MPC, we will focus only on the Chance-Constrained (CC) method here. The CC-MPC method was chosen for the statistical guarantees it provides, as well as its simplicity and similarity to MPC under specific assumptions on the uncertainties.
In CC-MPC and other stochastic MPCs, one change from MPC is the cost function. Before we minimized the actual cost of the system, but with the introduction of uncertainty, we will instead gain the optimum, by minimizing the expected cost of the system as seen in (\ref{eq:SMPCcost}). This allows for the cost function to be formulated by the same terms as we utilized in the MPC earlier.
\begin{equation}\label{eq:SMPCcost}
J = \min\limits_{u}E\bigg\{\sum\limits_{k=0}^N ||(\textbf{z}_k-\textbf{z}_k^{ref})||^2_Q+||\Delta \textbf{q}^u_k||^2_R\bigg\}
\end{equation}
By reformulation of the cost function into sums of expectations, the only non-constant quadratic term present is the left-hand side of (\ref{eq:exp}). While the right-hand side rewrites it to a quadratic term and a trace term of the variance of the output.
\begin{equation}
	E(\textbf{z}_k^TQ\textbf{z}_k) = E(\textbf{z}_k)^TQE(\textbf{z}_k) + tr(Q \sigma^2_{\textbf{z}_k}) \label{eq:exp}
\end{equation}
In addition to the reformulation of the cost function, the constraint set also needs to be reformulated. For CC-MPC, this is done in different ways for equality constraints and inequality constraints. The equality constraints are handled by taking the expectation of the constraints, similar to the cost function.

For the inequality constraints, a different approach is utilized. Here the deterministic constraints are handled on single constraints basis. Each constraint is rewritten as a probabilistic constraint as seen in (\ref{eq:SMPCcc}), which is a constraint on the probability of the deterministic constraint being true. The probability guaranty $\gamma$ ensures that the deterministic constraint is fulfilled for any rain inflow realization within the most likely $\gamma100\%$ realizations of the inflow. In (\ref{eq:SMPCcc}), the original inequality constraint is split in a stochastic part $f_{stoch}$ and in a deterministic part $f_{det}$.
\begin{equation}\label{eq:SMPCcc}
	Pr(f_{stoch} \leq f_{det}) \geq \gamma
\end{equation}
For CC-MPC, the stochastic part of the probabilistic constraint is assumed to follow a known distribution F, as given in (\ref{eq:stoch}).
\begin{equation}
	f_{stoch} \sim F(\bullet)\label{eq:stoch}
\end{equation}
Given that the probability function of constraint corresponds to the CDF of the deterministic part $\Phi_{F(\bullet)}(f_{det})$, the constraint can be rewritten using the quantile function of the distribution corresponding to $\gamma$, as below. 
\begin{equation}
	f_{det} \geq \Phi_{F(\bullet)}^{-1}(\gamma)
\end{equation}
Given that the quantile function above is depending on the parameters of the distribution, which likely are varying with the system, the constraint can be simplified by reformulation utilizing the CDF of a standardized distribution of F. In (\ref{eq:PC}), the standardized reformulation of distributions defined purely by their mean $\mu_f$ and their variance $\sigma_f^2$ is given, such as the normal distribution. For the remainder of this discussion, we will assume $f_{stoch}$ follows such a type of distribution, see (\ref{eq:stoch2}).
\begin{equation}
	\Phi_F\bigg(\frac{f_{det} - \mu_f}{\sigma_f}\bigg) \geq \gamma \label{eq:PC}
\end{equation}
\begin{equation}
	f_{stoch} \sim F(\mu_f, \sigma_f^2)\label{eq:stoch2}
\end{equation}
With the constraint being formulated with a standardized CDF, the quantile function $\Phi_F^{-1}$ of the standardized distribution F can again be applied to reformulate the constraints as (\ref{eq:PC1}), where only deterministic formulations are remaining. The usage of quantile functions is a potential drawback of CC-MPC, given that it requires that the quantile function of the given distribution to exist explicitly, or that the specific quantile corresponding to $\gamma$ is accessible for the computation.
\begin{equation}
	f_{det}  \geq \mu_f + \sigma_f\Phi_F^{-1}(\gamma) \label{eq:PC1}
\end{equation}
Doing the above reformulation for each inequality constraint and taking the expectation of the cost function and equality constraints, allows for the uncertainty to be removed from the formulation of the optimization problem. The inclusion of additional functions such as trace, mean and variance in the formulation, does potentially increase the complexity of the optimization. If some assumption on the uncertainty does hold, which is discussed below, the increased complexity can be avoided. Similarly discussed below is the case, when the distribution of the uncertainty was bounded by an interval. \\ \\

\subsubsection{CDF of Truncated Distribution}
\begin{figure}
      \centering
      \includegraphics[width = 0.5\textwidth,trim={0.8cm 0cm 1cm 0cm},clip ]{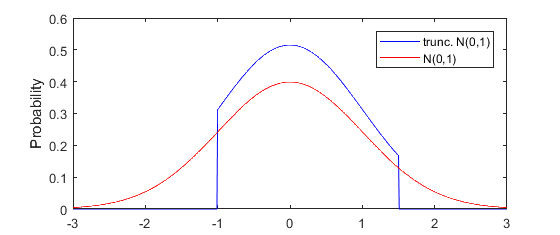}
      \caption{An illustration of a standard normal distribution, and a truncated standard normal distribution, truncated to within $-1$ and $1.5$}
      \label{fig:truncated}
\end{figure}

In some cases, the stochastic uncertainty is upper and/or lower bounded. Such a stochastic variable is said to follow a truncated distribution, as seen in (\ref{eq:TD}) for the variable X following a distribution F truncated between a and b. An illustration of the difference between an unbounded distribution and a truncated variant is given in Fig. \ref{fig:truncated}.
\begin{equation}
	X \sim F^b_a(\mu_x,\sigma_x^2) \Rightarrow  a\leq X \leq b\label{eq:TD}
\end{equation}
The CDF of a truncated distribution is given by (\ref{eq:CDFTD}). Where it can be seen to be formulated by the standardized unbounded distribution F. It can be observed that choosing either a or b to be infinity, indicating no bound, would remove the corresponding term in the CDF by either becoming 0 or 1 respectively or if both are chosen, becoming the standard distribution.
\begin{equation}
	\Phi_{F^b_a(\mu_x,\sigma_x^2)}(x) = \frac{\Phi_F(\frac{x-\mu_x}{\sigma_x}) - \Phi_F(\frac{a-\mu_x}{\sigma_x})}{\Phi_F(\frac{b-\mu_x}{\sigma_x}) - \Phi_F(\frac{a-\mu_x}{\sigma_x})}\label{eq:CDFTD}
\end{equation}
By rearranging and substitution of the truncated CDF, a bounded version of the probability constraint in (\ref{eq:PC}) can be formulated as (\ref{eq:BPC}). 
\begin{equation}
	\Phi_F(\frac{x-\mu_x}{\sigma_x}) \geq \gamma\Phi_F(\frac{b-\mu_x}{\sigma_x}) + (1-\gamma)\Phi_F(\frac{a-\mu_x}{\sigma_x})\label{eq:BPC}
\end{equation}
By applying the quantile function reformulation, the deterministic description of the probabilistic constraint with a bounded distribution can be formulated. \\ \\

\subsubsection{Assumptions on Uncertainty}
The CC-MPC is mathematically similar to MPC but does contain a few extra function terms in its formulation, such as the trace function, the quantile function and variances. All of these functions may potentially add to the computational difficulties for solving the optimization. However, if the following two assumptions hold, then the CC-MPC holds the same computational complexity as the deterministic MPC during the optimization.

The first assumption is on the independency of the uncertainty in the system, as stated:
\begin{itemize}
\item	the uncertain inflow of time k is independent of the uncertainty of the tank volume at time k, $\textbf{V}_k \perp \textbf{q}^{rain}_k$
\end{itemize}
This allows the variance $\sigma^2_{\textbf{z}_k}$ in (\ref{eq:exp}) to be rewritten as below, given that the covariance terms become zero.
\begin{equation}
	\sigma^2_{\textbf{z}_k} = C\sigma^2_{\textbf{V}_k}C^T + F\sigma^2_{\textbf{q}^{rain}_k}F^T
\end{equation}
The variance of the tank volume can with the same argumentation, be rewritten until it is only dependent on the initial variance of the volume and the variance of all inflows up until time $k-1$. This rewritten of the variance, leads to the trace term in (\ref{eq:exp}) becomes constant during an optimization. This reduces the cost function and the equality constraints to a deterministic formulation, where the variables correspond to the expected value instead of the actual value.

The second assumption is that the stochastic part of the constraints $f_{stoch}$ is independent of the optimization variables. This means that under an optimization, the mean and variance of the stochastic part is constant, resulting in the right-hand side of both (\ref{eq:PC1}) and (\ref{eq:BPC}) to also be constant. The resulting constraints have the same non-linearity as their MPC counterpart.

\section{Results \& Discussion}
In this section, the results of simulations with the different MPCs are discussed and compared. The MPC methods utilized are standard deterministic MPCs with perfect and imperfect predictions of the rain inflow, and CC-MPCs with different probability guarantees $\gamma$. For the MPCs, with imperfect predictions, 10 random realizations of the rain inflow were utilized for each simulation scenario, such that the realizations would be representative of the uncertainty. The probability guarantees $\gamma$ of the CC-MPC was utilized with all inequality constraints having the same probability with the following values being used: $95\%$, $90\%$, $80\%$, $70\%$, and $60\%$.

\begin{figure}
      \centering
     \includegraphics[width = 0.5\textwidth,trim={0.8cm 0cm 1cm 0cm},clip ]{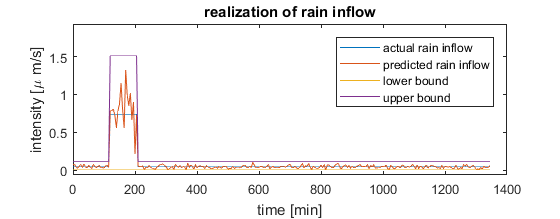}
    \caption{An example of actual rain inflow and a predicted rain inflow realization, shown for a step of $0.7$ $\mu m/s$ with a duration of $90$ minutes}
      \label{fig:rain}
\end{figure}

In the simulations, the rain inflow utilized were step inflows, which had varying rain durations and intensities. The rain steps utilized all had a dry weather period before and after the step with a duration of two and nineteen hours respectively. During the dry weather periods, the inflow to the sewers would be $0.04$ $\mu m/s$. The duration of the rain steps was varied with half-hour increments, starting at half an hour and ending at five-hour durations, while the intensity was varied from $0.1 \mu m/s$ to $11 \mu m/s$, with increments of $0.1 \mu m/s$. The inflow predictions utilized by the MPCs discussed above were generated as a truncated normal-distributed inflow with the mean being equal to the actual inflow, and the standard deviation $\sigma$ being $0.01$ $\mu m/s$ with a third of the actual inflow added to it. The lower bound on the truncated distribution was chosen to be zero, given that rain inflows are not negative. The upper bound was chosen to be three $\sigma$ above the expected inflow, corresponding to $99.7\%$ of rain inflow realizations for normal-distributed uncertainty. In Fig. \ref{fig:rain}, an example of the rain inflow is shown together with one realizations of the predicted rain inflow with its lower and upper bounds.

In the comparison, the MPC with perfect prediction will be utilized as a benchmark for the other MPCs. The focus of the comparisons is the avoidance of weir overflow generations. Given that the MPCs are designed to avoid overflows, there are only two scenarios for which an overflow would occur. The first scenario is the case for which an overflow is unavoidable, regardless of the control chosen; in other words, the optimization problem has become infeasible. The problem could have been infeasible from the start of the simulation (e.g. too heavy rain) or the previous choice of control has led it to be infeasible, due to uncertainty. In such cases, the controller relies on either previous computed predictions or other back-up procedures, such as less strict uncertainty handling. The second scenario is the case for which the MPC finds a feasible solution, but due to the uncertainties, the computed control leads to the generation of overflows. The feasible solution of this case is a false positive solution, in the sense that they are solution expected to be feasible, but are not. In the following sections, we will focus on each of these cases, but the scenario used for the comparison will only include those for which the MPC with perfect knowledge is feasible.

\subsection{Feasibility of Predictions}
\begin{figure}
      \centering
      \includegraphics[width = 0.5\textwidth,trim={0.8cm 0cm 1cm 0cm},clip ]{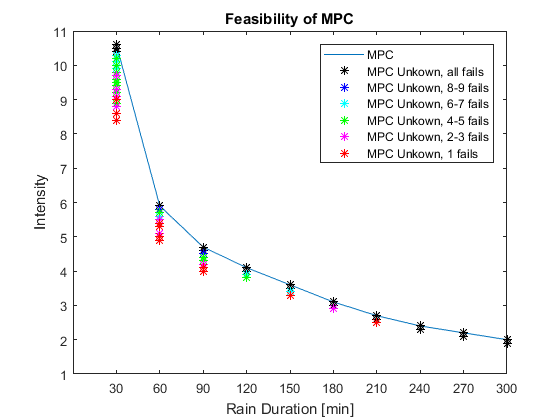}
      \caption{Feasibility of MPC with imperfect knowledge}
      \label{fig:MPCU:Feasbil}
\end{figure}

For the first case, the feasibility of the MPCs are the focus. In Fig. \ref{fig:MPCU:Feasbil}, we can observe the feasibility of the deterministic MPC with imperfect knowledge. Given that the blue line indicates the feasibility limit of the MPC with perfect knowledge, we can infer about the likelihood of the imperfect MPC being infeasible. We can observe that the closer the scenario is to the feasibility limits, the more of the ten realizations becomes infeasible. It is clear that for the scenarios with long duration goes directly from all realizations are feasible to all are infeasible.
While the scenarios with short durations slowly lose realizations to infeasibility as the intensity increases. Given that higher intensity in these simulations means higher variance, these results are in agreement with expectations of stochastic realizations of the uncertainty.

\begin{figure}
      \centering
      \includegraphics[width = 0.5\textwidth,trim={0.8cm 0cm 1cm 0cm},clip]{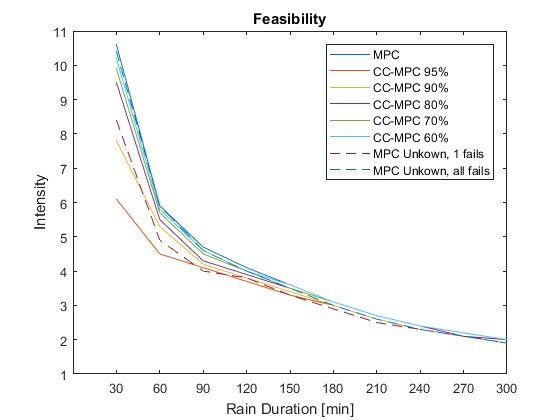}
      \caption{Feasibility of all three types of MPC}
      \label{fig:MPC:Feasbil}
\end{figure}

In Fig. \ref{fig:MPC:Feasbil}, the feasibility lines for the considered CC-MPCs are shown. It can clearly be seen that the CC-MPC with $\gamma=95\%$, stops being feasible significantly lower than the remainder of the CC-MPC or the perfect MPC. As the value of $\gamma$ decreases the feasibility lines close in on the MPC feasibility line. While higher values of $\gamma$  have more rain realizations covered by the solution, a slight decrease in value can restore feasibility, if the desired $\gamma$ would lead to infeasibility. This allows for a simple, but effective back-up strategy for CC-MPC; if the problem is infeasible, reduce the probability guarantee $\gamma$ until a feasible solution is found.

When we compare the CC-MPC feasibility lines with the colored feasibility points of the realizations of the imperfect MPC, we can observe that for a coverage of for example $\gamma=80\%$, the corresponding simulations with imperfect MPC have an infeasibility rate of around 4 out of 10. Giving approximately $40\%$ of rain realization not to be covered by the MPC, with additional scenarios below also having a higher chance of experiencing infeasibility. 
The discrepancy between the two types of MPCis most apparent, when the variance is higher, making the infeasibility tendencies converge as duration increases. This means that it is more likely that the use of CC-MPC will conserve feasibility, than by relying on a given rain prediction and a deterministic MPC. 

\subsection{False Positive Predictions}
\begin{figure}
      \centering
     \includegraphics[width = 0.5\textwidth,trim={5.5cm 1.5cm 4.5cm 3.5cm},clip ]{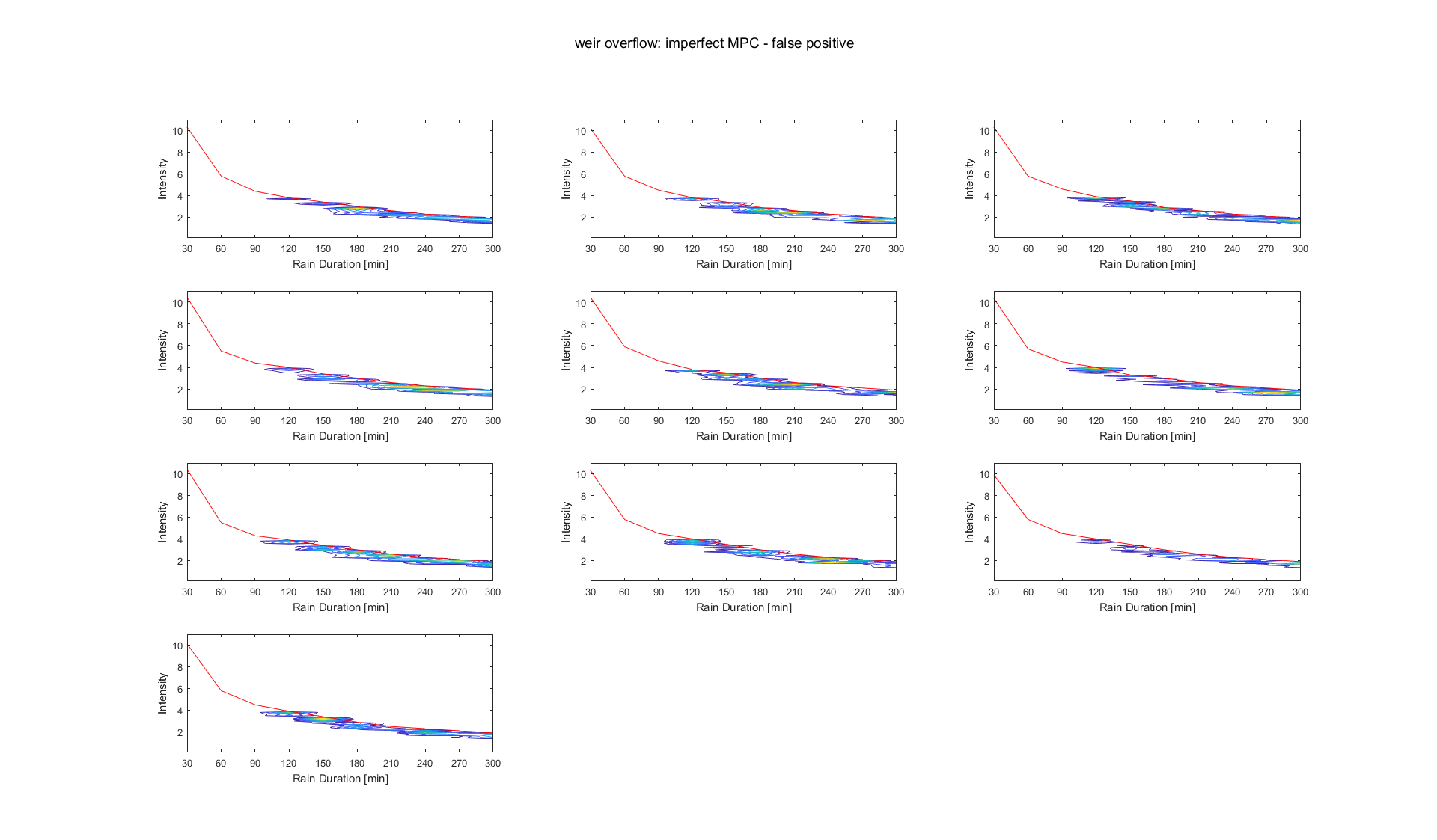}
    \caption{Weir overflow of MPC with imperfect knowledge, False positive predictions}
      \label{fig:MPCU:False}
\end{figure}

\begin{figure}
      \centering
     \includegraphics[width = 0.5\textwidth,trim={5.5cm 1.5cm 4.5cm 3.5cm},clip ]{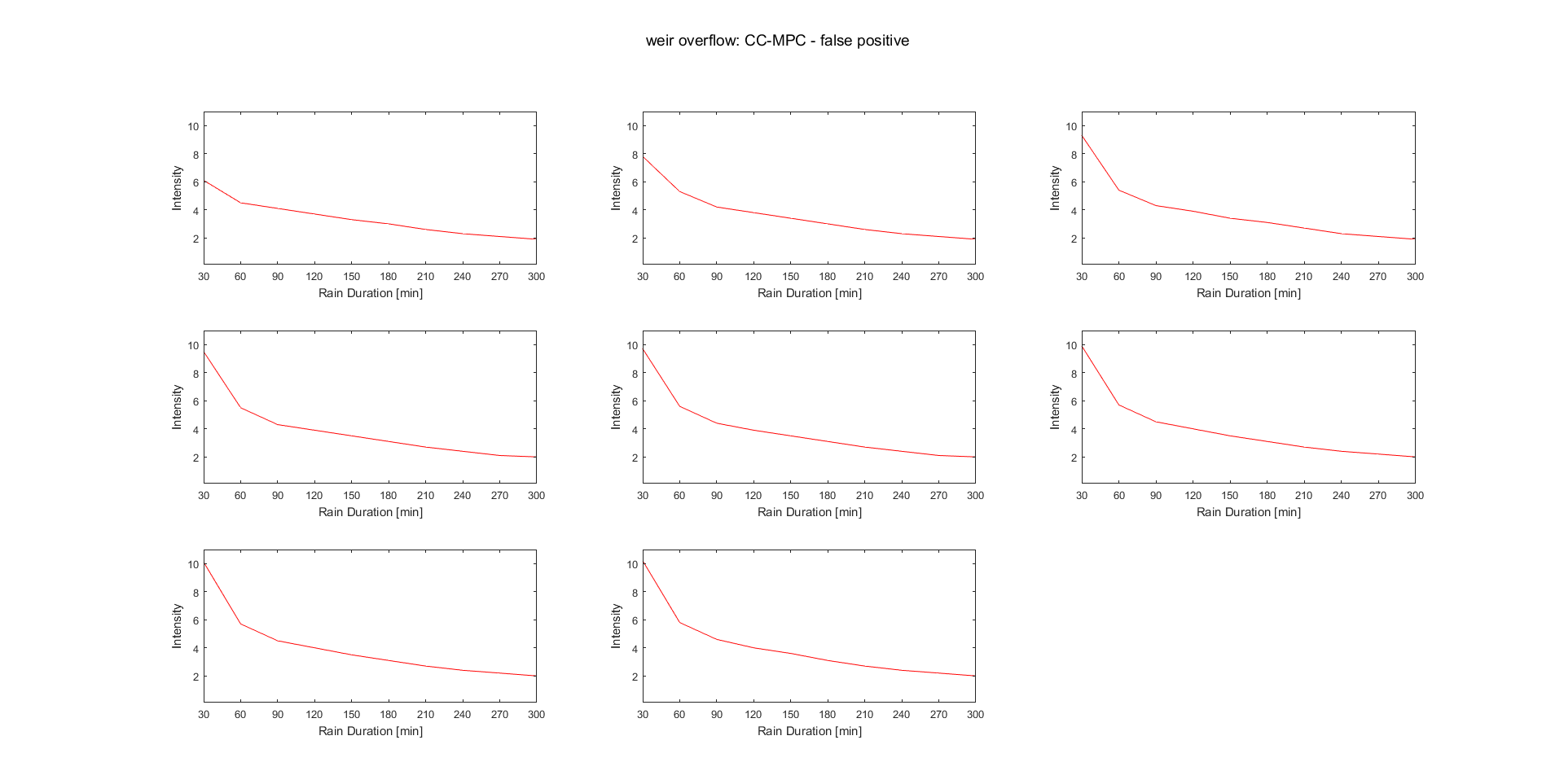}
    \caption{Weir overflow of Chance-Constrained MPC, False positive predictions}
      \label{fig:CCMPC:False}
\end{figure}

We will now focus on the case of false positive predictions, where the computed feasible predictions are in fact infeasible when implemented by the control. 
The occurrence of unpredicted weir overflows produced by the MPC with imperfect knowledge and by the CC-MPC can be seen in Fig. \ref{fig:MPCU:False} and Fig. \ref{fig:CCMPC:False} respectively. The red line in the figures is the feasibility line of the given MPC design. It can be observed that while the CC-MPC does not produce any false positive prediction of overflows, the basic MPC does.
We can further observe that the false positive overflow occur when the rain intensity gets closers to the feasibility line in the graph. 
It can further be seen that they only occur for longer rain durations, possible due to the increase in variance for the higher intensities at the lower durations; simply making the MPC infeasible. As is evident from the feasibility discussion of Fig. \ref{fig:MPCU:Feasbil}, showing how many realizations becoming infeasible at each scenario.

\subsection{Thoughts on different stochastic inflows}
In the previous sections, we have focused on stochastic forecasted rain inflows for which the actual inflow was equal to the expected inflow of the system, which is not generally the case. 
A simple deviation from this simple rain inflow is to assume the actual rain inflow has some bias, a sort of offset from the expected inflow of the system. When the bias makes the expected inflow larger than the actual inflow, it generally would lead to fewer overflows generated by both the MPC with imperfect knowledge and the CC-MPC, as long as the MPC's still is feasible. While a bias which makes the expected inflow lower than the actual inflow, generally would lead to more overflows and false positive predictions, due to the system being filled faster than expected. In the case of imperfect MPC, such bias would simply skew the probability of a given realization, therefore still leaving the feasibility of the MPC to chance. For the CC-MPC, a bias from the expected overflow would be accounted for by the constraint-restriction introduce by the probability guarantee $\gamma$, as long as the actual inflow does not correspond to a rain realization outside the covered realizations.

The rain considered in this work has been a simple step up and down; in reality, rains are a lot more varying and fluctuating in intensity. This means real rain has several spikes, and in general, an appearance closer to the uncertain predictions used earlier than actual steps.  On the other hand, given that MPC is a discrete method, any fluctuating rain can be described as a series of sample-length steps. This means that as long as the description of the stochastic distribution of the rain is known, then the CC-MPC can operate as seen if the actual rain lies within the bounds of a feasible probability guarantee $\gamma$.

%\section{Equipment}
%For the simulations performed in this work, the following hardware and software has been utilized
%\begin{itemize}
%\item MATLAB R2017a
%\item CVX v.2.1
%\item Mosek 8.1
%\item HP EliteBook Laptop with an Intel i7 processor
%\end{itemize}

\section{CONCLUSIONS}
In this paper, we have considered the utilization of the deterministic MPC and the stochastic Chance-Constrained MPC (CC-MPC) in a sewage system with uncertain predictions of the rain inflow.
From the results of the simulations, it was shown that CC-MPC was less prone to false positive predictions about feasibility and could keep the system feasible for a larger portion of possible rain realizations in comparison to the deterministic MPC. The CC-MPC being a more optimization-complex MPC type was shown to have the same complexity as the deterministic MPC, under the right assumptions on the uncertainty. In addition, the CC-MPC also had a clear and simple back-up procedure in case of infeasibility during computation, by slacking the desired probabilistic guarantees.


%%%%%%%%%%%%%%%%%%%%%%%%%%%%%%%%%%%%%%%%%%%%%%%%%%%%%%%%%%%%%%%%%%%%%%%%%%%%%%%%

%%%%%%%%%%%%%%%%%%%%%%%%%%%%%%%%%%%%%%%%%%%%%%%%%%%%%%%%%%%%%%%%%%%%%%%%%%%%%%%%

%%%%%%%%%%%%%%%%%%%%%%%%%%%%%%%%%%%%%%%%%%%%%%%%%%%%%%%%%%%%%%%%%%%%%%%%%%%%%%%%
%\appendix
%\section*{APPENDIX}

%\subsection{Mixed Logic Dynamics}\label{app:MLD}


%\section*{ACKNOWLEDGMENT}

%%%%%%%%%%%%%%%%%%%%%%%%%%%%%%%%%%%%%%%%%%%%%%%%%%%%%%%%%%%%%%%%%%%%%%%%%%%%%%%%

%References are important to the reader; therefore, each citation must be complete and correct. If at all possible, references should be commonly available publications.

\bibliographystyle{IEEEtran}
\bibliography{Sewer_CC-MPC_manuscript}

\end{document}